\documentclass[twocolumn,prl,showpacs,superscriptaddress]{revtex4-1}
\usepackage{amsmath}
\usepackage{amssymb}
\usepackage{graphicx}
\usepackage[dvipdfm,linkcolor=blue,anchorcolor=black,colorlinks=true,citecolor=blue]{hyperref}
\renewcommand{\section}[1]{{\par\it #1.---}\ignorespaces}

\begin{document}

\title{Mechanism for quantum speedup in open quantum systems}
\author{Hai-Bin Liu}
\affiliation{School of Physical Science and Technology, Lanzhou University, Lanzhou 730000, China}
\affiliation{School of Physics, Huazhong University of Science and Technology, Wuhan 430074, China}
\author{W. L. Yang}\email{ywl@wipm.ac.cn}
\affiliation{State Key Laboratory of Magnetic Resonance and Atomic and Molecular Physics, Wuhan Institute of Physics and Mathematics, Chinese Academy of Sciences, Wuhan 430071, China}
\author{Jun-Hong An}\email{anjhong@lzu.edu.cn}
\affiliation{School of Physical Science and Technology, Lanzhou University, Lanzhou 730000, China}
\author{Zhen-Yu Xu}
\affiliation{School of Physical Science and Technology, Soochow University, Suzhou, 215006, China}

\begin{abstract}
The quantum speed limit (QSL) time for open system characterizes the most efficient response of the system to the environmental influences. Previous results showed that the non-Markovianity governs the quantum speedup. Via studying the dynamics of a dissipative two-level system, we reveal that the non-Markovian effect is only the dynamical way of the quantum speedup, while the formation of the system-environment bound states is the essential reason for the quantum speedup. Our attribution of the quantum speedup to the energy-spectrum character can supply another vital path for experiments when the quantum speedup shows up without any dynamical calculations. The potential experimental observation of our quantum speedup mechanism in the circuit QED system is discussed. Our results may be of both theoretical and experimental interests in exploring the ultimate QSL in realistic environments, and may open new perspectives for devising active quantum speedup devices.
\end{abstract}
\maketitle

\section{Introduction}
As one of the fundamental laws of nature, quantum mechanics imposes a bound on the evolution speed to quantum systems, the so-called quantum speed limit (QSL) \cite{QSL1,QSL2,QSL3,QSL4,QSL5}. It has recently attracted considerable attention and played remarkable roles in various areas of quantum physics including nonequilibrium thermodynamics \cite{Deffner1}, quantum metrology \cite{Metro1,Metro2,Metro3,Metro4}, quantum optimal control \cite{Con1,Con2,Con3,Con4,Con5,Con6}, quantum computation \cite{QC1,QC2,QC3}, and quantum communication \cite{QSL1,Qcom}. The QSL time sets a bound on the minimal time a system needs to evolve between two distinguishable states, and it can be understood as a generalization of the time-energy uncertainty principle. For isolated systems, the QSL time under unitary evolution is determined by the maximum \cite{Lev} of the Mandelstam-Tamm bound $\tau_\text{MT}=\pi \hbar /(2\Delta E)$ \cite{QSL2,QSL3} and Margolus-Levitin bound $\tau _\text{ML}=\pi \hbar /(2 \bar{E})$ \cite{ML}, where $\Delta E$ and $\bar{E}$ are the fluctuation and mean value of the initial-state energy, respectively.

Because of the inevitable interactions with the environments, quantum systems should be generally regarded as open systems. The decoherence effect resulting from the system-environment interactions would introduce remarkable influences on the QSL. Much effort has been made to explore the QSL of open system under the environment governed nonunitary evolution. A Mandelstam-Tamm-type bound on the QSL time for pure initial states has been derived by using positive nonunitary maps \cite{Tadd,Cam}. Using a geometric approach, a unified bound on the QSL time including both Mandelstam-Tamm and Margolus-Levitin types has also been formulated \cite{Deffner2}. The generic bound on the QSL time for both mixed and pure initial states has been obtained by introducing the relative purity \cite{YJZ} and the Hilbert-Schmidt product of operators \cite{Fidelity,ZSun} as distance measure. Deffner and Lutz \cite{Deffner2} showed that non-Markovian effect characterized by non-Markovianity \cite{Wolf2008,Breu,Rivas,Luo} can speed up quantum evolution, which has been verified in different settings \cite{YJZ,ZSun,ZYX1,ZYX2}. A first experimental observation in cavity QED systems \cite{Exp}, where an atomic beam is treated as a controllable environment for the cavity field system, has confirmed such non-Markovianity-assisted speedup. Despite the growing body of literature on this subject, the analysis has almost exclusively been focused on the direct relation between the quantum speedup and the non-Markovianity. To our knowledge, no ultimate physical explanation has been provided for the mechanism of quantum speedup in open systems to date.

Here we give a constructive answer that both the quantum speedup phenomenon and the non-Markovianity are attributed to the formation of the system-environment bound states (as special eigenstates with eigenvalues residing in the band gap of the energy spectrum) \cite{bound1,bound2,bound3,An1,An2}. Via considering explicitly a two-level system (TLS) coupled dissipatively to a bosonic environment, we reveal that the formation of the bound states significantly changes the dynamics of the TLS, and thus plays a critical role in quantum speedup and the non-Markovian dynamics. Possible experimental realization for our quantum speedup mechanism is discussed. Our result suggests that one can control the QSL of open systems via engineering the formation of the bound state.

\section{The model and dynamics}
Consider a TLS coupled to a bosonic environment, a widely used model in studying the QSL of open systems. Its Hamiltonian is ($\hbar=1$)
\begin{equation}
\hat{H}=\omega _{0}\hat{\sigma}_{+}\hat{\sigma}_{-}+\sum_{k}\omega _{k}\hat{a}_{k}^{\dagger }\hat{a}_{k}+\sum_{k}(g_{k}\hat{a}_{k}^{\dagger }\hat{\sigma}_{-}+\text{H.c.}), \label{hamlilt}
\end{equation}
where $\hat{\sigma}_{\pm }$ and $\omega _{0}$ are the transition operators and frequency of the TLS, and $\hat{a}_{k}^{\dagger }$ ($\hat{a}_{k}$) is the creation (annihilation) operator of the $k$-th environmental mode with frequency $\omega _{k}$. The coupling strength between the TLS and its environment is denoted by $g_{k}$, which is further characterized by the spectral density $J(\omega )=\sum_{k}\left\vert g_{k}\right\vert^{2}\delta (\omega -\omega _{k})$.

The non-Markovian dynamics of the TLS is governed by the exact master equation
\begin{eqnarray}
\dot{\rho}(t)&=&-i{\Omega(t)\over 2}[\hat{\sigma}_{+}\hat{\sigma}_{-},\rho(t)]+{\Gamma(t)\over 2}[2\hat{\sigma}_{-}\rho(t)\hat{\sigma}_{+}\nonumber\\
&&-\hat{\sigma}_{+}\hat{\sigma}_{-}\rho(t)-\rho(t)\hat{\sigma}_{+}\hat{\sigma}_{-}]\equiv\check{\mathcal{L}}_{t}\rho (t), \label{mastereq}
\end{eqnarray}
where $\Gamma(t)+i\Omega(t)\equiv-2\dot{c}(t)/c(t)$ with $c(t)$ determined by
\begin{equation}
\dot{c}(t)+i\omega _{0}c(t)+\int_{0}^{t}f(t-t' )c(t' )dt' =0,
\label{eq:c-t}
\end{equation}
under the condition $c(0)=1$. Here $f(x)=\int_0^\infty J(\omega)\exp (-i\omega x)d\omega$ is the environmental correlation function. The time-dependent $\Omega(t)$ and $\Gamma(t)$ are the renormalized frequency and the decay rate, respectively. All the non-Markovian memory effect has been registered self-consistently in these time-dependent coefficients. Note that the decay rate $\Gamma(t)$ is by no means positively definite. Actually it is just the negative $\Gamma(t)$ which causes the energy backflow from the environment to the TLS. It is seen as a main non-Markovian feature. In the Markovian limit, one can prove $c(t)\simeq\exp[-i(\omega_0-\delta\omega)t-\pi J(\omega_0)t]$ with $\delta\omega=\mathcal{P}\int_0^\infty {J(\omega)d\omega\over\omega-\omega_0}$, which reduces $\Omega(t)\simeq 2(\omega_0-\delta\omega)$ and $\Gamma(t)\simeq 2\pi J(\omega_0)$. This positive constant decay rate ensures the energy to flow unidirectionally from the TLS to the environment.

\section{QSL and non-Markovianity}
QSL for open systems characterizes how fast a system under the environmental driving can evolve. Hence, via evaluating the characteristic of the QSL, one could obtain the efficient response of the system to the environmental disturbance. The QSL time between an initial state $\rho (0)=|\psi _{0}\rangle \langle \psi _{0}|$ and its target state $\rho(\tau)$ for open systems is defined by $\tau _\text{QSL}=\sin^{2} \mathcal{B}[\rho (0),\rho (\tau )]/\Lambda_\tau^{(\infty)}$ \cite{Deffner2}, where $\mathcal{B}[ \rho (0),\rho (\tau )] \equiv\arccos \sqrt{\langle\psi _{0}\vert \rho (\tau )\vert \psi _{0}\rangle }$ is the Bures angle between $\rho (0) $ and $\rho (\tau )$, and $\Lambda_{\tau }^{(\infty)}=( 1/\tau )\int_{0}^{\tau }dt||\check{\mathcal{L}}_{t}\rho (t)|| _{\infty}$ with the operator norm $\Vert \hat{A}\Vert_{\infty}$ equaling to the largest eigenvalue of $\sqrt{\hat{A}^\dag \hat{A}}$. From Eq. \eqref{mastereq}, the QSL time for our model
\begin{equation}
\tau _\text{QSL}=\frac{1-|c(\tau)|^{2}}{(1/\tau )\int_{0}^{\tau }| \partial _{t}|c(t)|^{2}| dt},  \label{eq:QSL}
\end{equation} is achieved when $|\psi_0\rangle=|+\rangle$.
Once $c(t)$ has been calculated from Eq. \eqref{eq:c-t}, $\tau _\text{QSL}$ could be deduced by Eq. \eqref{eq:QSL}.

The non-Markovian effect can be quantified by the non-Markovianity $\mathcal{N}=\max_{\rho _{1,2}(0)}\int_{\sigma >0}dt\sigma (t,\rho _{1,2}(0))$, where $\sigma (t,\rho_{1,2}(0))=\dot{D}(\rho _{1}(t),\rho _{2}(t))$ is the change rate of the trace distance $D(\rho _{1}(t),\rho _{2}(t))=\frac{1}{2}\text{Tr}\vert \rho _{1}(t)-\rho_{2}(t)\vert $ between states $\rho_{1,2}(t)$ evolving from their respective initial forms $\rho_{1,2}(0)$ \cite{Breu}. In the Markovian limit, the environment acts like a ``sink" such that all the energy flows irreversibly from the system to the environment. The states become more and more indistinguishable with time evolution, which means $\sigma<0$ always and thus $\mathcal{N}=0$. In the non-Markovian dynamics, the dynamical interplay between the system and the environment would transiently cause the energy backflow from the environment to the system. This would lead to the increase of the state distance. Hence the non-Markovianity is defined to characterize the non-Markovian effect just via quantifying such environmental backaction induced increase of the distinguishability of the given states. For our TLS, it has been proven that the optimal pair of initial states to maximize $\mathcal{N}$ are $\rho_{1,2}(0)=|\pm\rangle\langle\pm|$ \cite{Deffner2}, which means $D(\rho _{1}(t),\rho _{2}(t))=|c(t)|^2$. Then one can readily verify
\begin{equation}
\mathcal{N}={1\over 2}\big[|c(\tau)|^2-1+\int_0^\tau |\partial_t|c(t)|^2|dt\big],
\end{equation}which connects to $\tau_\text{QSL}$ as $\tau_\text{QSL}=\tau{1-|c(\tau)|^2\over 1-|c(\tau)|^2+2\mathcal{N} }$ \cite{ZYX1}.

\section{The mechanism of quantum speedup}
It has been found that the non-Markovian effect can speed up quantum evolution \cite{Deffner2,ZYX1}. As another dynamical quantity, the non-Markovianity could not be seen as an essential reflection to the QSL of the dynamical evolution. On the other hand, such attribution of the quantum speedup to the non-Markovianity is experimentally meaningless because the non-Markovianity is not an experimental observable. The question is: what is the essential mechanism of quantum speedup in a more physical angle of view? We here try to provide an ultimate explanation to the physical mechanism of quantum speedup.

\begin{figure}[tbp]
\includegraphics[width=.95\columnwidth]{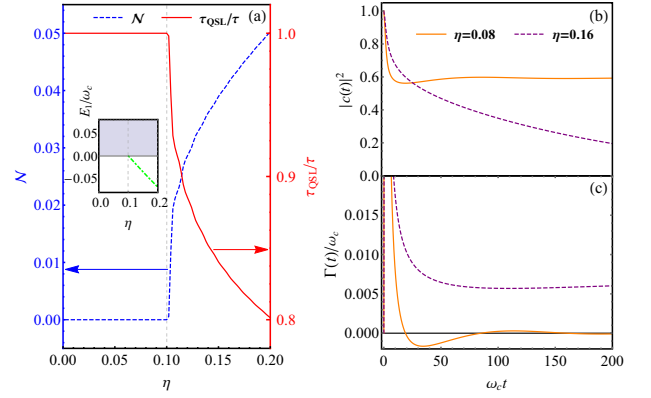}
\caption{(Color online) (a) Non-Markovianity $\mathcal{N}$ (blue dashed line) and QSL time $\tau_\text{QSL}$ (red solid line) as a function of the coupling constant $\eta$. The inset shows the energy spectrum of the total system, where the green dot-dashed line denotes the energy of the formed bound state. The obtained $|c(t)|^2$ (b) and the corresponding decay rate $\Gamma(t)$ (c) with and without bound state. The bound state is formed when $\eta >\eta _{c}=0.1$. The parameters are $\omega_{c}\tau=800$, $\omega _{0}/\omega _{c}=0.1$, and $s=1$.}\label{fig:1}
\end{figure}

Different from the widely studied Lorentzian spectrum \cite{Deffner2,YJZ,ZSun,ZYX1,ZYX2}, we here study the Ohmic kind of spectra with the spectral density $J(\omega )=\eta \omega ^{s}\omega _{c}^{1-s}e^{-\omega /\omega _{c}}$, where $\eta $ is a coupling constant, $\omega _{c}$ is a cutoff frequency, and $s$ is a power index of the spectrum. It can be classified as Ohmic $(s=1)$, sub-Ohmic case $(0<s<1)$, and super-Ohmic case $(s>1)$ \cite{Leg}. This kind of spectra can be simulated in an ion trap system \cite{Porras2008}. We plot in Fig. \ref{fig:1}(a) the non-Markovianity $\mathcal{N}$ and the QSL time $\tau_\text{QSL}$ as a function of $\eta$ for $s=1$. Here we choose the final time $\tau$ sufficiently large. This not only guarantees the convergence of the calculated $\mathcal{N}$, which, determined by the system parameters, is a quantity independent on $\tau$, but also ensures that the system reaches its steady state and thus the obtained $\tau_\text{QSL}$ reflects the evolution efficiency of the system relaxing to its equilibrium state. It is interesting to find that both $\tau_\text{QSL}$ and $\mathcal{N}$ show an obvious threshold at the critical point $\eta_c=0.1$. When $\eta <\eta _{c}$, $\tau _\text{QSL}/\tau $ stays at $1$ and $\mathcal{N}$ remains zero. Thus there does not exist any quantum speedup and non-Markovian effect. However, when $\eta\geq\eta_c$, $\tau_\text{QSL}/\tau$ experiences a steep decrease and $\mathcal{N}$ experiences a steep increase. It indicates that the system presents a dramatic speedup and non-Markovian effect. Although this result confirms the previous result that the quantum speedup connects directly with the non-Markovianity \cite{Deffner2,ZSun,ZYX1}, the existence of the obvious threshold point $\eta_c$ inspires us to further pursue the hidden physical reason for the quantum speedup.

Our main point is that the non-Markovian dynamics of an open system connects tightly with the energy-spectrum characters of the total system consisting of the system and its environment. Thus the study of the energy spectrum may supply us with a meaningful message to understand its dynamics.
Since $\hat{N}=\hat{\sigma}_+\hat{\sigma}_-+\sum_k\hat{a}_k^\dag\hat{a}_k$ is conserved, the Hilbert space splits into independent subspaces with definite $N$. For our zero-temperature environment case, only the subspaces with $N=0$ and $1$ are involved in the dynamics. Besides the trivial eigenstate $|\varphi_0\rangle=|-,\{0_k\}\rangle$ with $E_0=0$ for the $N=0$ subspace, we can obtain the eigenstate of the $N=1$ subspace as $\left\vert \varphi_1\right\rangle =d_{0}\left\vert +,\{0_{k}\}\right\rangle+\sum_{k}d_{k}\left\vert -,1_{k}\right\rangle $ with $d_0=[ 1+\int \frac{J(\omega )}{\left( E_{1}-\omega \right) ^{2}}d\omega ] ^{-1/2}$ and $E_1$ satisfying \cite{Miy}
\begin{equation}
y(E_1)\equiv\omega _{0}-\int_{0}^{\infty }\frac{J(\omega )}{\omega -E_1}d\omega =E_1.
\label{eq:bound-state}
\end{equation}
Since $y(E_1) $ decreases monotonically with the increase of $E_1$ in $E_1<0$ and $\lim_{E_1\rightarrow -\infty}y(E_1 )=\omega _{0}$, Eq. \eqref{eq:bound-state} has an isolated root in the bandgap [see Fig. \ref{fig:1}(a)] whenever
\begin{equation}
y(0)<0\Rightarrow \eta >\frac{\omega _{0}}{\omega _{c}}\frac{1}{\gamma (s)}= \eta _{c}
\label{eq:critical-value}
\end{equation}with $\gamma(s)$ the gamma function. We call the eigenstate corresponding to this isolated eigenvalue $E_{\text{BS}}$ the bound state. Note that the negativity of  $E_{\text{BS}}$ for our TLS cannot cause the unboundedness from below in a harmonic oscillator system \cite{Paz2015}. It is because no canonical transformation can make the combination $\hat{C}=\alpha\hat{\sigma}_-+\sum_k \beta_k\hat{a}_k$ fulfill $[\hat{C},\hat{C}^\dag]=1$. Thus $n E_{\text{BS}}$ with $n$ an arbitrary integer by no means is also the eigenvalue of the TLS. It is remarkable to find from Fig. \ref{fig:1}(a) that the critical point for forming the bound state matches well with the one for presenting quantum speedup. We thus conjecture that the formation of the bound state of the total system is the essential mechanism of the quantum speedup of the open TLS.

The distinguished role played by the bound state in the quantum speedup can be understood in the following way. Consider the initial state $|\Psi(0)\rangle=|+,\{0_k\}\rangle$, whose evolution can be expanded as
\begin{equation}
|\Psi(t)\rangle=q_0e^{-iE_{\text{BS}}t}|\varphi_{1,\text{BS}}\rangle+\sum_{\alpha\in\text{CB}}q_\alpha e^{-iE_{1,\alpha}t}|\varphi_{1,\alpha}\rangle, \label{expan}
\end{equation}
where $|\varphi_{1,\text{BS}}\rangle$ is the potentially formed bound state with the eigenenergy $E_{\text{BS}}$, $|\varphi_{1,\alpha}\rangle$ are the eigenstates in the continuous energy band with eigenenergies $E_{1,\alpha}$, $q_0=\langle \varphi_{1,\text{BS}}|\Psi(0)\rangle=d_0$, and $q_\alpha=\langle \varphi_{1,\alpha}|\Psi(0)\rangle$. Due to the out-of-phase interference contributed by the continuous energies $E_{1,\alpha}$, all the excited-state population in the components of the summation in Eq. \eqref{expan} tends to vanish and only the one in the bound-state component survives in the long-time limit, i.e.,
\begin{equation}
\text{Tr}[\hat{\sigma}_+\hat{\sigma}_-\rho(\infty)]=|c(\infty)|^2=d_0^4,\label{cstd}
\end{equation}where the first equality is obtained from Eq. (\ref{mastereq}). If the bound state is absent, then $|c(\infty)|^2$ approaches zero asymptotically. Figures \ref{fig:1}(b) and \ref{fig:1}(c) plot the excited-state population $|c(t)|^2$ and the decay rate $\Gamma(t)$, respectively. We can see that $\Gamma(t)$ in the absence of the bound state when $\eta<\eta_c$ tends to a positive constant after a short-time jolt. The complete positivity of $\Gamma(t)$ causes $|c(t)|^2$ to decay to zero monotonically. Here the environment has no backaction on the system and thus $\mathcal{N}$ is zero. Since the TLS equilibrates in an asymptotic manner to its ground state, there is no more efficient evolution than the evolution characterized by $\tau$ and thus $\tau_\text{QSL}=\tau$. When the bound state is formed in the condition $\eta >\eta _{c}$, the competition between the environmental backaction and the dissipation effects on the TLS causes $\Gamma(t)$ to transiently take negative value and asymptotically approaches zero [see Fig. \ref{fig:1}(c)]. Consequently, after some short-time oscillations, $|c(t)|^2$ tends to a finite value matching well with the result in Eq. (\ref{cstd}). It is contrary to one's expectation that a larger coupling strength always induces a stronger decoherence. The transient increase of $|c(t)|^2$ causes the increase of the distinguishability and the decrease of $\tau_\text{QSL}$. The system here has a great capacity to speed up. Hence the bound state supplies a latent capability, while the non-Markovianity only supplies the dynamical way to the system for quantum speedup.

\begin{figure}[tbp]
\includegraphics[width=.95\columnwidth]{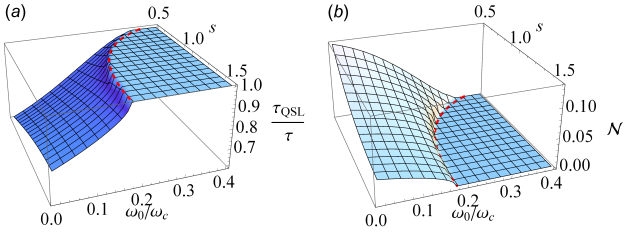}
\caption{(Color online) (a) QSL time $\tau_\text{QSL}$ and (b) non-Markovianity $\mathcal{N}$ as a function of the spectral power index $s$ and the system frequency $\omega_{0}$. The red dashed line shows the critical values for forming the bound state. The parameters are $\omega _{c}\tau=800$ and $\eta =0.2$.}\label{fig:2}
\end{figure}

The deterministic role of the bound state in the quantum speedup is expected to also be valid for other spectra and systems. Figure \ref{fig:2} shows $\tau_\text{QSL}$ and $\mathcal{N}$ with the change of the system frequency and the spectral index. The results confirm that the threshold point from Markovian to non-Markovian and from no-speedup to speedup regimes matches well with the one for forming the bound state. Our mechanism is applicable to a dissipative harmonic oscillator, whose dynamics is governed by the same bound-state mechanism as our TLS \cite{Lu2013, Yang2014}, and the spin-boson model, where the bound state has been found playing a leading role in the quantum phase transition of the model \cite{Tong2011, An1}. Thus it is generic in open systems. Note that our mechanism cannot be captured by the previous treatment with Lorentzian spectrum \cite{Deffner2,YJZ,ZSun,ZYX1,ZYX2}, where the lower limit of the frequency integration in $f(t-\tau)$ is artificially extended from $0$ to $-\infty$. This approximation is mathematically convenient but loses the availability of the bound state \cite{An2}.

\section{Physical realization}
\begin{figure}[tbp]
\includegraphics[width=.95\columnwidth]{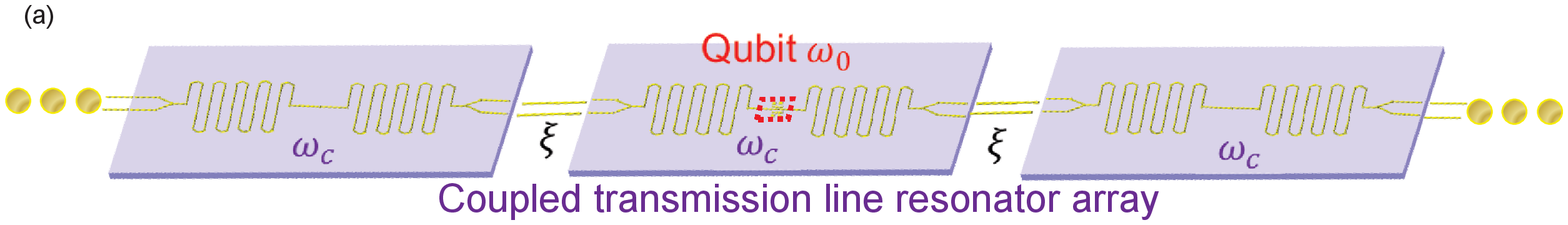}\\
\vspace{.3cm}
\includegraphics[width=.95\columnwidth]{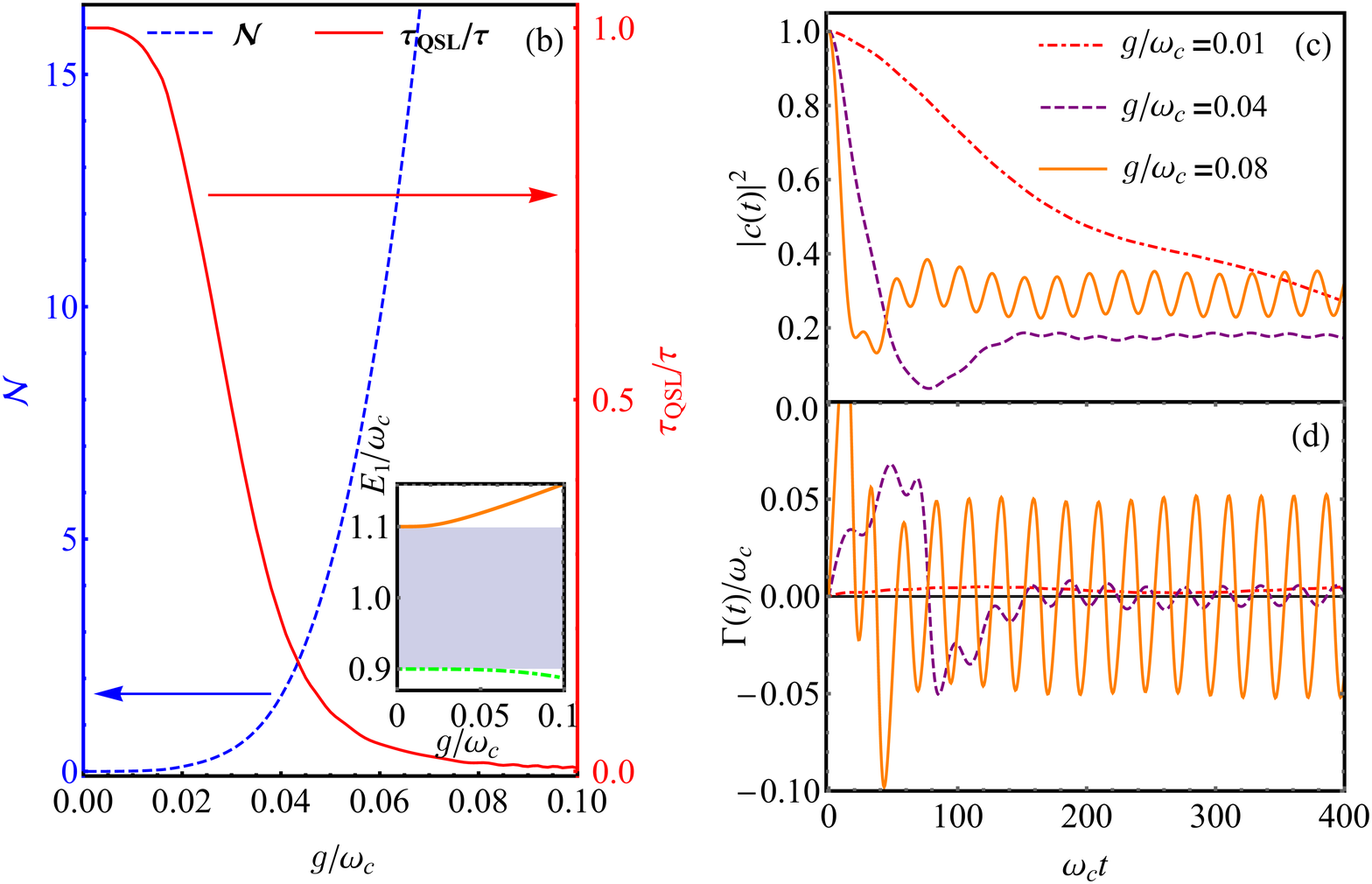}
\caption{(Color online) (a) Suggested proposal of a superconducting qubit interacting with an array of coupled transmission line resonators. (b) Non-Markovianity $\mathcal{N}$ and QSL time $\tau_\text{QSL}$ as a function of coupling strength $g$. The inset shows the energy spectrum of the total system, where the green dot-dashed and the the orange lines denote the energy of the formed bound state. The obtained $|c(t)|^2$ (c) and the corresponding decay rate $\Gamma(t)$ (d) with and without bound state. The parameters are $\xi =0.05\omega _{c}$, $\omega _{0}=1.08\omega _{c}$, and $N=800$.}\label{fig:3}
\end{figure}

The ideal physical system to verify our prediction is a superconducting qubit interacting with an array of coupled microwave superconducting resonators \cite{Matinis2011,cir1,cir2,cir3} [see Fig. \ref{fig:3}(a)] governed by $\hat{H}=\omega _{0}\hat{\sigma}_{+}\hat{\sigma}_{-}+\omega _{c}\sum_{j=-N/2}^{N/2}\hat{b}_{j}^{\dagger }\hat{b}_{j}+\xi\sum_{j=-N/2}^{N/2-1}(\hat{b}_{j+1}^{\dagger }\hat{b}_{j}+g\hat{\sigma}_{+}\hat{b}_{0}+\text{H.c.})$. It becomes
\begin{equation}
\hat{H}=\omega _{0}\hat{\sigma}_{+}\hat{\sigma}_{-}+\sum_{k}\epsilon _{k}\hat{b}_{k}^{\dagger }\hat{b}_{k}+\frac{g}{\sqrt{N}}\sum_{k}( \hat{\sigma}_{+}\hat{b}_{k}+\text{H.c.})
\end{equation}
through the Fourier transform $\hat{b}_{j}=\sum_{k}\hat{b}_{k}e^{ikjx_{0}}$, where $\epsilon _{k}=\omega _{c}+2\xi \cos kx_{0}$, and $x_{0}$ is the spatial separation of the two neighbor resonators. Here the coupled resonator array acts as a synthesized environment with finite bandwidth. The novel character of this system is that two bound states at most could be formed, which causes
\begin{equation}
|c(\infty)|^2=d_{0,1}^4+d_{0,2}^4+2d_{0,1}^2d_{0,2}^2\cos[(E_{\text{BS}1}-E_{\text{BS}2})t]
\end{equation}with $d_{0,j}=[ 1+\int \frac{J(\omega )}{\left( E_{\text{BS}j}-\omega \right) ^{2}}d\omega ] ^{-1/2}$. The equilibrium state of the TLS here shows persistent oscillation between the two bound states. It is expected that $\mathcal{N}$ cannot converge with respect to the final time $\tau$ and $\tau_\text{QSL}$ tends to zero. The features of $\tau_\text{QSL}$, $\mathcal{N}$, $|c(t)|^2$, and $\Gamma(t)$ are plotted in Figs. \ref{fig:3}(b)-\ref{fig:3}(d). When $\eta \lesssim 0.015$, no bound state is formed and no speedup is present. When $0.015\lesssim \eta \lesssim 0.07$, one bound state is formed, and there are remarkable quantum speedup and non-Markovianity. When $\eta \geq 0.07$, two bound states are formed, their effect is greatly enhanced and $\tau _\text{QSL}/\tau $ almost reduces to zero and $\mathcal{N}$ tends to divergence. These results are in agreement with the above theoretical predictions.

Our result implies that one can control the QSL time through engineering the bound state via manipulating the environmental spectrum.  This idea is physically similar to the results in Ref. \cite{Exp}, where the quantum speedup is experimentally observed by controlling the environment. The decoherence suppression induced by the bound state has been observed in the photonic crystal system \cite{Lodahl2004, Noda2007, Dreisow08} and hopefully is realizable in the ion trap system for the Ohmic-type spectra \cite{Porras2008}. Furthermore, the experimental advances in the array of coupled resonators offer unprecedented potential of engineering dynamical behavior of quantum systems \cite{cir1,cir2}. The recent experiments have demonstrated an excellent quantum control on the multi-resonator interconnected by phase qubits or other superconducting elements \cite{Matinis2011}. Therefore, our proposed scheme is experimentally accessible.

\section{Conclusions}
In summary, we have explored the mechanism of quantum speedup in a dissipative TLS. It is revealed that the bound state of the whole system, as an isolated eigenstate with eigenvalue residing in the bandgap of the eigenspectrum, significantly changes the decoherence dynamics and plays a decisive role in quantum speedup. This mechanism can supply experimenters with a useful guideline to judge when the quantum speedup shows up without any calculations to its explicit dynamics. We have also proposed a scheme to verify our prediction. Addressing the realization of quantum speedup in realistic open quantum systems, our work opens a new avenue to control the QSL via engineering the formation of the bound state.

\section{Acknowledgments}
This work is supported by the Specialized Research Fund for the Doctoral Program of Higher Education, by the Program for New Century Excellent Talents in University, and by the National Natural Science Foundation of China (Grants No. 11175072, No. 11274351, No. 11204196, and No. 11474139).

\end{document}